\documentclass[12pt,preprint]{aastex}
\usepackage{epsfig,wrapfig,graphicx}

\def\ms{m~s$^{-1}$}
\def\ks{km~s$^{-1}$}
\def\msini{$M_P\sin{i}$}

\def\vsini{$V_{\rm rot}\sin{i}$}
\def\msun{$M_{\odot}$}
\def\mjup{$M_{\rm Jup}$}
\def\rsun{$R_{\odot}$}
\def\lsun{$L_{\odot}$}
\def\chisq{$\sqrt{\chi^2_\nu}$}

\def\feh{[Fe/H]}
\def\rphk{$R^\prime_{HK}$}
\def\shk{$S$}

\def\npllet{three}
\def\nplcap{Three}

\def\pC{351.5}
\def\peC{6}
\def\pB{297.3}
\def\peB{6}
\def\pA{341.1}
\def\peA{7}
\def\tpC{2452994}
\def\tpeC{30}
\def\tpB{2450213}
\def\tpeB{20}
\def\tpA{2453118}
\def\tpeA{40}
\def\eC{0.149}
\def\eeC{0.06}
\def\eB{0.33}
\def\eeB{0.2}
\def\eA{0.152}
\def\eeA{0.08}
\def\omC{54}
\def\omeC{30}
\def\omB{183}
\def\omeB{30}
\def\omA{301}
\def\omeA{30}
\def\kC{51.3}
\def\keC{5}
\def\kB{14.0}
\def\keB{2}
\def\kA{39.2}
\def\keA{4}
\def\rmsC{9.2}
\def\rmsB{5.6}
\def\rmsA{7.4}
\def\chiC{1.14}
\def\chiB{1.01}
\def\chiA{1.07}
\def\nobsC{34}
\def\nobsB{29}
\def\nobsA{29}
\def\mstarC{1.68}
\def\mstarB{1.65}
\def\mstarA{1.85}
\def\msiniC{2.5}
\def\msiniB{0.61}
\def\msiniA{2.0}
\def\arelC{1.16}
\def\arelB{1.03}
\def\arelA{1.17}

\def\starA{HD\,210702}
\def\starB{HD\,175541}
\def\starC{HD\,192699}
\def\hippA{HIP\,109577}
\def\hippB{HIP\,92895}
\def\hippC{HIP\,99894}
\def\mstarA{1.85}
\def\mstarB{1.65}
\def\mstarC{1.68}
\def\rstarA{4.72}
\def\rstarB{3.85}
\def\rstarC{4.25}
\def\lstarA{13.1}
\def\lstarB{9.56}
\def\lstarC{11.5}
\def\bvA{0.951}
\def\bvB{0.869}
\def\bvC{0.867}
\def\mvA{2.19}
\def\mvB{2.49}
\def\mvC{2.30}
\def\vmagA{5.93}
\def\vmagB{8.03}
\def\vmagC{6.44}
\def\ageA{1.4}
\def\ageB{1.9}
\def\ageC{1.8}
\def\dA{56}
\def\dB{128}
\def\dC{67}
\def\shkA{0.11}
\def\shkB{0.11}
\def\shkC{0.12}
\def\rhkA{-5.35}
\def\rhkB{-5.28}
\def\rhkC{-5.29}

\def\feA{$+0.12$}
\def\feB{$-0.07$}
\def\feC{$-0.15$}
\def\loggA{3.29}
\def\loggB{3.52}
\def\loggC{3.44}
\def\teffA{5010}
\def\teffB{5060}
\def\teffC{5220}
\def\vsiniA{1.7}
\def\vsiniB{2.9}
\def\vsiniC{1.9}

\begin{document}
\title{Retired A Stars and Their Companions: Exoplanets
  Orbiting \nplcap\ Intermediate--Mass Subgiants$^1$}

\author{ John Asher Johnson\altaffilmark{2}, 
  Debra	A. Fischer\altaffilmark{3}, 
  Geoffrey W. Marcy\altaffilmark{2},
  Jason T. Wright\altaffilmark{2},
  Peter Driscoll\altaffilmark{4},
  R. Paul Butler\altaffilmark{5},
  Saskia Hekker\altaffilmark{6},
  Sabine Reffert\altaffilmark{7},
  Steven S. Vogt\altaffilmark{8}
}

\email{johnjohn@astron.berkeley.edu}

\altaffiltext{1}{Based on observations obtained at the Lick
  Observatory, which is operated by the University of California, and
  W. M. Keck Observatory, which is operated jointly by the University
  of California and the California Institute of Technology}
\altaffiltext{2}{Department of Astronomy, University of California,
Mail Code 3411, Berkeley, CA 94720}
\altaffiltext{3}{Department of Physics \& Astronomy, San Francisco
  State University, San Francisco, CA 94132}
\altaffiltext{4}{Department of Earth \& Planetary Sciences, Johns
  Hopkins University, Baltimore, MD, 21218} 
\altaffiltext{5}{Department of Terrestrial Magnetism, Carnegie
  Institution of Washington DC, 5241 Broad Branch Rd. NW, Washington DC,
 20015-1305}
\altaffiltext{6}{Leiden Observatory, Leiden University, PO Box 9513,
  2300 RA Leiden, The Netherlands}
\altaffiltext{7}{ZAH-Landessternwarte, K\"onigstuhl 12, 69117
  Heidelberg, Germany}
\altaffiltext{8}{UCO/Lick Observatory, University of California at
  Santa Cruz, Santa Cruz, CA 95064}

\begin{abstract}
We report precision Doppler measurements of \npllet\
intermediate--mass subgiants obtained at Lick and Keck
Observatories. All \npllet\ stars show variability in
their radial velocities consistent with planet--mass companions in
Keplerian orbits. We find a planet with a minimum mass
$M_P\sin{i} = \msiniC$~\mjup\ in a \pC~day orbit around \starC, a planet
with a minimum mass of \msiniA~\mjup\ in a \pA~day orbit around
\starA, and a planet with a minimum mass of \msiniB~\mjup\ in a
\pB~day orbit around \starB. Mass estimates from stellar interior 
models indicate that all three stars were formerly A--type,
main--sequence dwarfs with masses ranging from \mstarB~\msun\ to
\mstarA~\msun. These \npllet\ long--period planets would not have been
detectable during their stars' main--sequence phases due to the large
rotational velocities and stellar jitter exhibited by early--type
dwarfs. There are now 9 ``retired'' (evolved) A--type 
stars ($M_* >  1.6$~\msun) with known planets. All 9 planets orbit at
distances $a \geq 0.78$ ~AU, which is significantly different than the
semimajor axis 
distribution of planets around lower--mass stars. We examine the
possibility that the observed lack of short--period planets is due to
engulfment by their expanding host stars, but we find that this
explanation is inadequate given the 
relatively small stellar radii of K giants ($R_* < 32$~\rsun~$=0.15$~AU) and 
subgiants ($R_* < 7$~\rsun~$ = 0.03$~AU). Instead, we conclude that planets 
around intermediate--mass stars reside preferentially beyond
$\sim$0.8~AU, which may be a reflection of different formation and
migration histories of planets around A--type stars.   
\end{abstract}


\section{Introduction}

Very little is known about the occurrence rate and orbital properties  
of planets around A--type stars, corresponding to stellar masses
ranging from 1.6~\msun\ to 3.0~\msun.  
Inspection of the Catalog of Nearby Exoplanets (CNE)\footnote{For the
  updated catalog of extrasolar planet and their parameters see
  http://exoplanets.org.} reveals that only 6 of the 173 stars with
securely detected planetary companions have masses in excess of
1.6~\msun \citep{butler06}. This small number of detections
is not a true reflection of the occurrence of planets
around A--type stars, but rather the result of a strong selection bias
against early--type, main--sequence stars in precision Doppler surveys.

Measuring precise Doppler shifts of early--type dwarfs is complicated 
by their rotationally broadened spectral features, high surface
temperatures, and high levels of excess radial velocity noise, or
``jitter'' \citep{saar98, wright05}. \citet{galland05} find that Doppler
precision for early--type dwarfs is limited to $\sim 40$~\ms\ at
spectral type F5V, and 90--200~\ms\ for A stars, rendering Doppler
measurements of these stars sensitive only to planets with large
masses and short orbital periods. The lowest mass companion 
so far detected around an A star is the brown
dwarf orbiting HD\,180777 \citep{galland06}. Even though the 
28~day orbital solution has a large velocity semiamplitude, $K
=1200$~\ms, the signal is only a 3$\sigma$ detection above the stellar
jitter and measurement uncertainties. 

Most of what is known about planet formation around intermediate--mass
stars comes from two primary sources: direct imaging of disks around
young stars and Doppler detections of planets around evolved stars.
While A--type dwarfs are poor Doppler targets, their
high intrinsic luminosities facilitate the detection and direct
imaging of material in their circumstellar environments. 
More than a decade before the discoveries of the first extrasolar
planets, evidence of planet formation outside of our Solar
System came from the infrared detection of collision--generated
dust around the A--type, main sequence stars Vega \citep{aumann84} and
$\beta$~Pic \citep{smith84}. Since then, advances in high--contrast
imaging have resulted in the detection of an optically thick disk around a
pre--main--sequence Herbig Ae star \citep{perrin06}, as
well as scattered light images of optically thin "debris disks" around
11 main--sequence stars---the majority of which have spectral types F5V
or earlier \citep[Table 2 of][and references therein]{kalas06,
  schneider06, wahhaj07}. Recent observations of the 
debris disk around the young A star Fomalhaut have revealed a
perturbation in the disk structure that may be due to the influence of
an orbiting Jovian planet \citep{kalas05}. Studying the relationships
between the architectures of disks around young A stars and the
distribution of planet properties around their older counterparts will
provide key tests of planet formation models.

A key to finding planets around A stars using Doppler methods is
provided by the effects of 
stellar evolution. As stars evolve away from the main sequence, they
become cooler and rotate slower, which increases the number of narrow
absorption lines in their spectra \citep{gray85, schrijver93, 
 donascimento00}. 
Several Doppler surveys have focused on evolved, intermediate--mass
stars on the red giant branch \citep{frink02, mitchell04, hatzes05,
  lovis05} and 
clump giant branch \citep{sato03, setiawan03}. These surveys have resulted in
the discovery of 6 substellar companions orbiting
former A--type stars (Table \ref{massive_table}). That none of these
planets would have been detectable during their host stars'
main--sequence phases highlights the important role evolved stars play
in the study of planets around intermediate--mass stars.  

Here we present \npllet\ new planet candidates around stars with $M_*
 > 1.6$~\msun. These detections come from our precision Doppler survey
 of evolved stars on the subgiant branch of the H--R diagram. We
discussed the selection criteria of our target stars in
\citet{johnson06b}, along with the discovery of an eccentric hot Jupiter
orbiting the 1.28~\msun\ subgiant HD\,185269. We discuss
our spectroscopic observations and Doppler measurement technique in
\S~\ref{observations}. In \S~\ref{stars}, we present the 
characteristics of the host stars along with the orbital solutions
for their planet candidates. We conclude with a comparison of the
semimajor axis distributions of planets around A--type stars
and lower--mass stars in \S~\ref{summary}.    

\section{Observations}
\label{observations}

We are monitoring a sample of 159 evolved stars at Lick and Keck
Observatories \citep{johnson06b}. At Lick Observatory, the Shane 3\,m
and 0.6\,m 
Coude Auxiliary Telescopes (CAT) feed the Hamilton spectrometer
\citep{vogt87}, which has a resolution of $R \approx 50,000$ at
$\lambda = 5500$~\AA. Spectroscopic observations at Keck Observatory were
obtained using the HIRES spectrometer with a resolution of $R \approx
80,000$ at $\lambda = 5500$~\AA\ \citep{vogt94}.  
Doppler shifts are measured from each spectrum
using the iodine cell method described by \citet{butler96} \citep[see
  also][]{marcy92b}. A temperature--controlled Pyrex cell containing
gaseous iodine is placed at the entrance slit of the
spectrometer. The dense set of narrow molecular lines imprinted on
each stellar spectrum from 5000 to 6000~\AA\ provides a
robust wavelength scale for each observation, as well as information
about the shape of the spectrometer's instrumental response. 

\begin{figure}[t!]
\epsscale{0.8}
\plotone{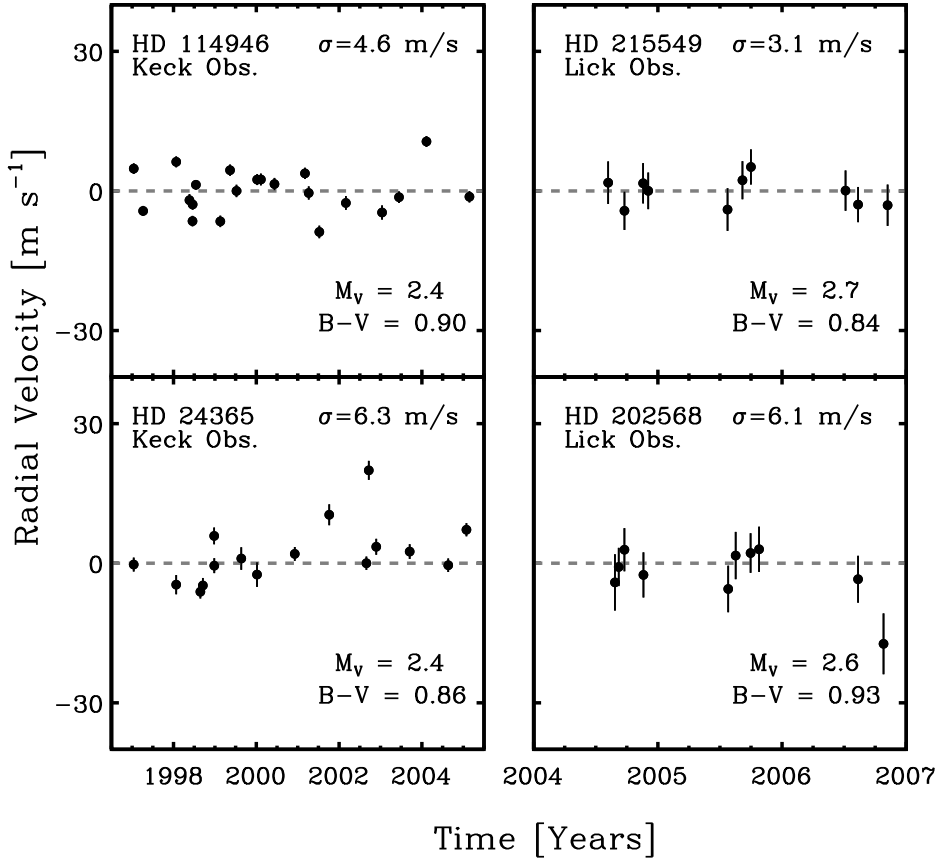}
\caption {\footnotesize{Radial velocity time series for four stable subgiants with
  $B-V$ colors and absolute visual magnitudes similar to those of the
  \npllet\ subgiant planet host stars. $\sigma$ represents the
  standard deviation of the velocities about the mean (dashed
  lines). Observations 
  from Keck (left   panels) and Lick (right panels) show that
  subgiants in this region of the H--R diagram are typically stable
  to well within 10~\ms\ over time scales of many years. \label{std_stars}}}  
\end{figure}

Traditionally, the Doppler shift of each stellar observation is made
with respect to an observed, iodine--free stellar template spectrum. 
These template observations require higher signal and resolution than 
normal radial velocity observations, which leads to increased exposure
times. Given our large target list and the small aperture of the CAT,
obtaining an observed template for each star would represent a
prohibitive cost in observing time. We therefore perform
a preliminary analysis of each star's observations using a
synthetic, ``morphed'' template spectrum following the method
described by \citet{johnson06}. Stars showing conspicuous Doppler
variations are reanalyzed using a traditional, observed template to
verify the signal and search for a full orbital solution. 

Doppler measurements from Keck and Lick Observatories for four stable
subgiants are shown in Figure~\ref{std_stars}. The error bars
represent the internal uncertainties 
of each measurement, which are approximated by the weighted standard
deviation of the mean velocity measured from each of the 700 individual
2~\AA\ wide chunks in each spectrum \citep{butler96}.  We
typically achieve internal 
measurement uncertainties of 1-2~\ms\ for Keck observations and
3-5~\ms\ at Lick. Subgiants have an additional 4-6~\ms\ of
``jitter''---velocity scatter in excess of internal errors due to
astrophysical sources such as pulsation and rotational modulation of
surface features \citep[][]{saar98, wright05}. We therefore adopt a
jitter value of 5~\ms\ for our subgiants, which is added in quadrature
to the internal uncertainties of the measurements before
searching for a best--fit orbital solution. 

After determining the best--fit Keplerian solution using a
Levenberg--Marquardt, least--squares minimization, we estimate the 
orbital parameter uncertainties using a bootstrap 
Monte Carlo method. We first subtract the best--fit Keplerian from the
measured velocities. The residuals are then scrambled and added back
to the original measurements, and a new set of orbital parameters is
obtained. This process is repeated for 1000 trials, and the standard
deviations of the parameters from all trials are adopted as
the formal, 1$\sigma$ uncertainties.  

\section{Stellar Properties and Orbit Solutions}
\label{stars}

\subsection{Estimates of Stellar Properties}
\label{stellar}

We estimated the stellar properties of our target stars using two
primary methods: the LTE
spectral synthesis method (SME) described by \citet{valenti05}, and
the Padova\footnote{See also http://pleiadi.pd.astro.it/} stellar
interior models. The spectral synthesis method uses a 
non--linear least--squares algorithm to vary the parameters of a
synthetic spectrum to search for a fit to an iodine--free stellar
template spectrum. The free parameters in the fit are the 
abundances of heavy elements; effective surface temperature, $T_{\rm eff}$;
surface gravity, $\log{g}$; and broadening effects due to
the star's projected rotation velocity, \vsini. \citet{valenti05}
estimate a precision of 0.04 dex in 
metallicity, 44~K in effective temperature, 0.3 dex in $\log{g}$, and
0.5~\ks\ in rotational velocity.

\begin{figure}[t!]
\epsscale{0.8}
\plotone{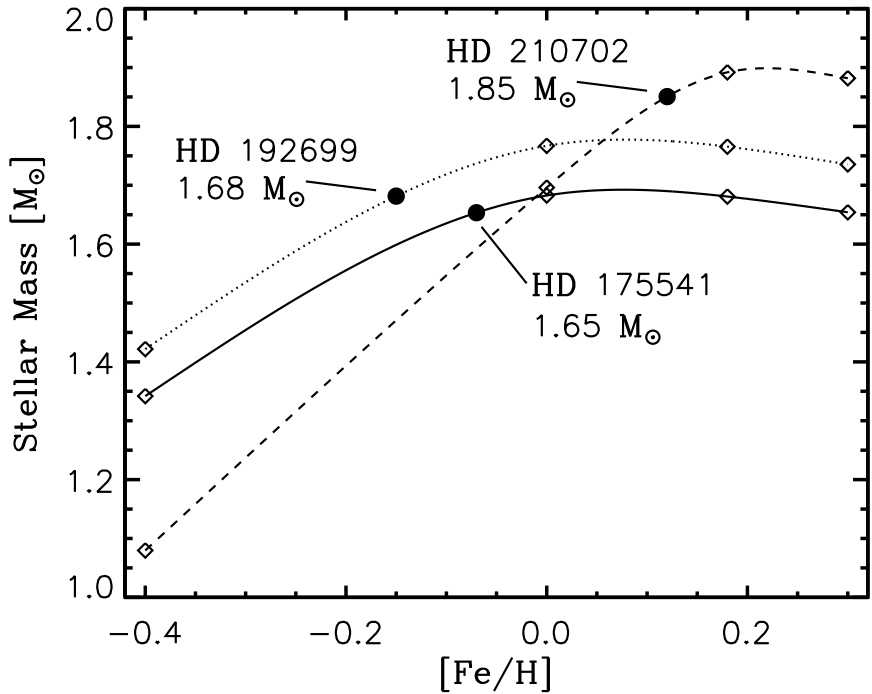}
\caption{\footnotesize{This figure illustrates the interpolation method employed to 
  determine accurate stellar masses for the planet host stars. We
  estimated each star's mass, radius and age by interpolating its
  Hipparcos $B-V$ 
  color and absolute visual magnitude $M_V$ onto grids of four
  different metallicities: [Fe/H] $= -0.4, 0.0, +0.18, +0.3$ (open
  diamonds). For each star's measured value of $B-V$ and $M_V$, the
  three lines show the dependence of stellar mass on the measured
  [Fe/H], estimated using a cubic spline interpolation between the
  diamonds. Similar dependencies were determined for stellar radii,
  luminosities and ages. \label{massint}}    }
\end{figure}

To estimate stellar masses, radii, luminosities and ages, we used the Padova
theoretical stellar models, which have been 
transformed into several photometric systems by
\citet{girardi02}. Stellar properties can be inferred by interpolating
a star's color, absolute magnitude and metallicity onto these model
grids. However, the \citet{girardi02} model grids are defined at
widely--spaced metallicity intervals, with [Fe/H]~$ = $~-0.4, 0.0,
+0.18 and +0.30. Since the uncertainties in our spectroscopically
derived metallicity estimates are much less than the model grid
intervals, and because the derived stellar properties do not vary
linearly with [Fe/H], we could not simply perform a linear, 3-dimensional
interpolation of $M_V$, $B-V$ and [Fe/H]. Instead, we first linearly
interpolate the stars' colors and absolute magnitudes onto each of the 
four metallicity grids. We then use a cubic spline interpolation
between the grid points to measure the desired stellar property
(e.g. mass) at the star's measured [Fe/H].
Our procedure is illustrated in Figure~\ref{massint}, which
shows stellar mass as a function of [Fe/H] for each star's
absolute magnitude and color. The same procedure was used for stellar
radii, luminosities and ages.

We compared our interpolated stellar properties to the \citet{takeda06}
theoretical interior models of the stars in the Spectroscopic
Properties of Cool Stars catalog 
\citep[SPOCS][]{valenti05}. We found a subset of 11 evolved stars 
in the catalog with $2.0 < M_V < 3.0$ and $0.7 < B-V < 1.1$.
Differences between our inferred values and those from \citet{takeda06}
had an rms scatter of 7\% in mass, 12\% 
in radius, with a median offset of -2\% and -4\% in each parameter,
respectively. Ages of this subset of evolved stars estimated by 
the two methods have a difference
of -0.4~Gyr with and rms scatter of 1.1~Gyr. We therefore adopt
fractional uncertainties of 7\% for our derived masses, 12\% for
radii and 1~Gyr for ages. We list the full set of derived stellar
properties of the three candidate planet host stars in
Table~\ref{stellartable}. We summarize each star's properties and 
orbital solution in the following subsections.  

\subsection{\starC}

\starC\ (\hippC) is listed with a G5 spectral type in the \emph{Hipparcos}
Catalog, with $V = \vmagC$, $B - V = \bvC$ and a parallax--based
distance of \dC~pc \citep{hipp}. However, no luminosity class is
given. Based on its distance, we calculate $M_V = \mvC$, which at its
$B-V$ color places the star 3.7 mag above the mean main--sequence of stars
in the Solar neighborhood, as defined by \citet{wright04}. Based on
its color and absolute magnitude, we find that \starC\ is
likely a G8\,IV subgiant near the base of the red giant
branch. Commensurate with its evolved status, 
\starC\ is chromospherically inactive, with \shk~$=$~\shkC\ and
\rphk~$=$~\rhkC\ as measured from the CaII H\&K line core and averaged
over all observations \citep{wright04b}. 

Based on our LTE spectral analysis, we find that \starC\ is 
metal--poor, with $\rm [Fe/H] = -0.15$, and slowly rotating, with
\vsini~$ = \vsiniC$~\ks. The other stellar parameters derived
from our spectral analysis are listed in Table~\ref{stellartable}. We
interpolated the star's color, absolute magnitude and metallicity
onto the \citet{girardi02} theoretical stellar model grids using the
method described 
in \S~\ref{stellar}. Our interpolation yields a stellar mass $M_* =
\mstarC$~\msun, radius $R_* = \rstarC$~\rsun, and an age of \ageC~Gyr. 

\begin{figure}[t!]
\epsscale{1}
\plotone{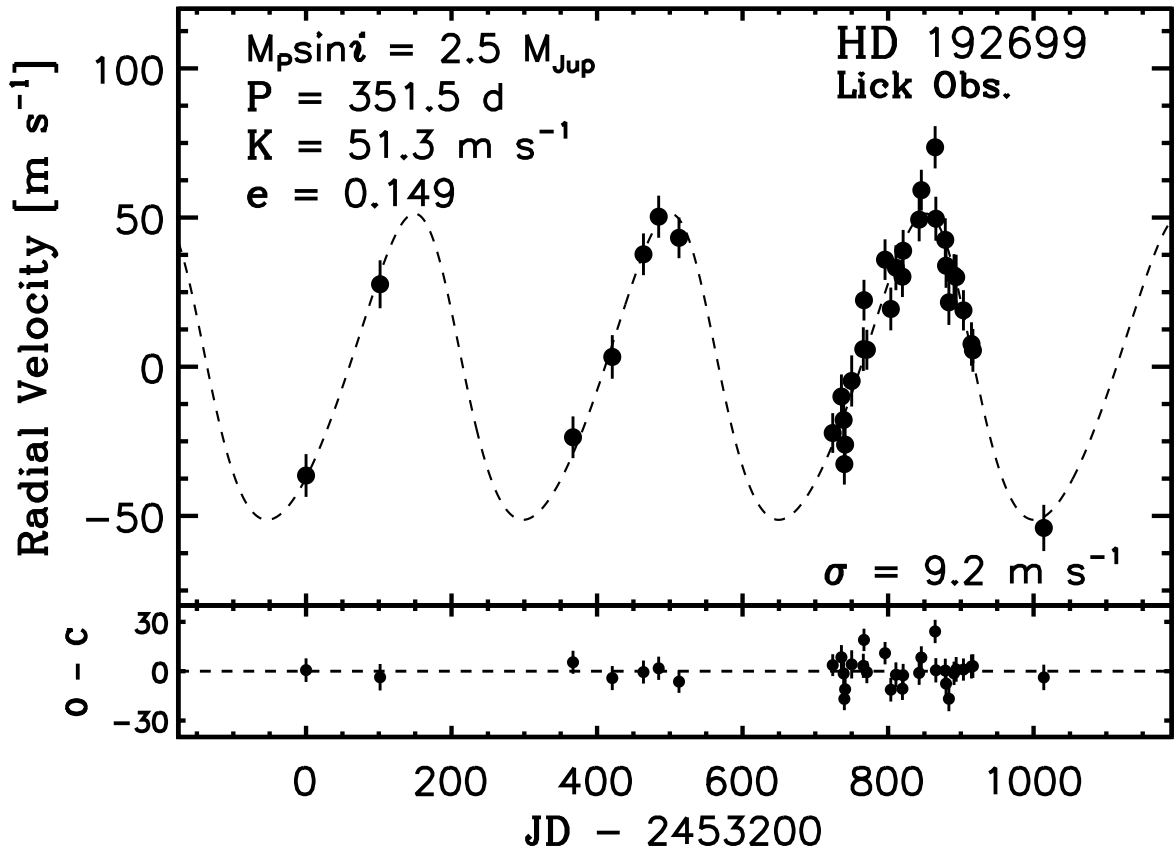}
\caption {\footnotesize{Radial velocity time series for \starC\
  measured at Lick 
  Observatory. The error bars reflect the quadrature sum of the
  internal measurement uncertainties and 5~\ms\ of jitter. The dashed
  line shows the   best--fit orbital solution, which has \chisq$ =
  \chiC$. \label{orbitC}}  }
\end{figure}

We began observing \starC\ in 2004 May at Lick Observatory using the
3~m Shane Telescope and 0.6~m CAT.
Table~\ref{vel192699} lists our \nobsC\ velocity measurements, along with
their times of observation and internal measurement uncertainties
(without jitter). Our
first 7 observations, initially analyzed using a synthetic stellar template
spectrum \citep{johnson06}, showed correlated variations spanning
two observing seasons. We obtained a high--quality observed template
using the Shane~3m telescope and initiated intensive follow--up
observations during the Fall 2006 observing season. The Keplerian signal
is visible to the eye (Figure \ref{orbitC}), obviating a periodogram
analysis.  

The best--fit
Keplerian orbit has a period of $P = \pC$~d, velocity amplitude $K =
\kC$~\ms, and eccentricity $e = \eC \pm \eeC$. With an assumed stellar
mass of \mstarC~\msun, we estimate a minimum planet mass \msini~$ =
\msiniC$~\mjup\ and orbital separation $a = \arelC$~AU. The fit has
$\rm rms = \rmsC$~\ms\ and a reduced \chisq~$= \chiC$, consistent with
the measurement errors and jitter. The full set of orbital parameters
and uncertainties is listed in Table~\ref{orbittable}.

\subsection{\starA}
\label{starA}

\starA\ (\hippA, HR\,8461) is listed in the Hipparcos catalog as a K1
star (no luminosity class given) with $V = \vmagA$, $B-V = \bvA$, and a
parallax--based distance of \dA~pc. Given its distance and apparent
magnitude, we calculate an absolute magnitude $M_V = \mvA$, which
places it 4.2~mag above the average main sequence of stars in the
Solar neighborhood \citep{wright04}. We therefore estimate that
\starA\ is a class K1\,IV subgiant near the base of the red giant branch.

Based on our LTE spectral analysis, we find that \starA\ is somewhat
metal--rich, with \feh~$=\feA \pm 0.04$, and slowly rotating, with \vsini~$ = \vsiniC$~\ks. Our 
interpolation of the star's color, absolute magnitude and metallicity
onto the \citet{girardi02} stellar model grids yields a stellar mass $M_* =
\mstarA$~\msun, stellar radius $R_* = \rstarA$~\rsun, and an age of
\ageA~Gyr. Consistent with its post--main--sequence evolutionary
status, \starA\ is chromospherically inactive with \shk~$=$~\shkA\ and 
\rphk~$=$~\rhkA, as measured from its CaII~H\&K emission
\citep{wright04b}.  The other stellar parameters derived
from our spectral analysis and stellar model interpolation are listed
in Table~\ref{stellartable}. 

\begin{figure}[h!]
\epsscale{1}
\plotone{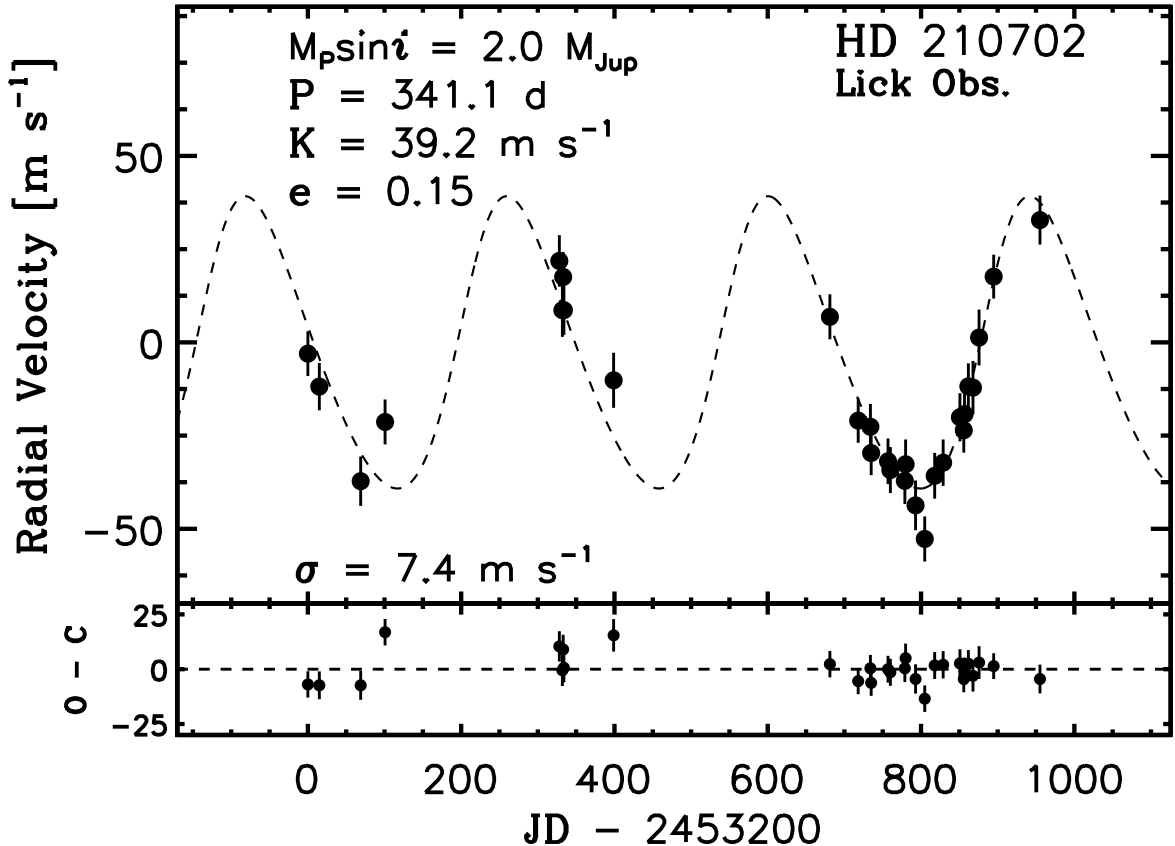}
\caption {\footnotesize{Radial velocity time series for \starA\
  measured at Lick  Observatory. The dashed line shows the
  best--fit orbital solution, which has \chisq$=\chiA$.\label{orbitA}} }
\end{figure}

We began monitoring \starA\ in 2004 August at Lick Observatory. The
first 9 observations were Doppler--analyzed using a synthetic
template, and showed excessive variability with $\rm
rms=19$~\ms. We then obtained a traditional, observed template
to confirm the variations with higher Doppler precision. The full set
of velocities is listed in Table~\ref{vel210702} (without jitter) and
plotted in Figure \ref{orbitA}. The error bars in Figure~\ref{orbitA}
have been augmented by adding 5~\ms\ of  
jitter in quadrature to the internal measurement uncertainties. 

The best--fit Keplerian orbital solution is shown in
Figure~\ref{orbitA} overplotted on the velocities. The solution has a
\pA~day period, an eccentricity  $e = \eA \pm \eeA$, and a
semiamplitude $K = \kA$~\ms. The fit residuals have $\rm rms =
\rmsA$~\ms\ and reduced \chisq~$=$~\chiA, consistent with the
internal measurement uncertainties and jitter. Assuming a stellar mass
$M_* = \mstarA$~\msun, the best--fit solution yields a relative separation $a
= \arelA$~AU. 

We find that the inclusion of a linear trend in the orbital solution
yields a slight improvement in the quality of fit, decreasing the rms
scatter of the residuals from \rmsA~\ms\ to 6.7~\ms, and the reduced
\chisq\ from \chiA\ to 1.00 after accounting for the
extra free parameter in the 
Keplerian--plus--trend model. We tested the validity of the trend
using the prescription of \citet{wright07}, and found a false--alarm
probability of 49\%. The large FAP indicates that the apparent
linear trend is likely due to noise rather than and additional orbital
companion. Indeed, the trend appears to be driven primarily by the
three outliers near JD~$=$ 100, 400 and 800 (Figure~\ref{orbitA}). We
therefore favor the single--planet Keplerian model summarized in
Table~\ref{orbittable}. 

\subsection{\starB}
\label{sectionB}

\starB\ (\hippB) is listed in the
\emph{Hipparcos} Catalog as a G8V star with $V = \vmagB$, $B-V = \bvB$
and a parallax--based distance of \dB~pc \citep{hipp}. Given its
distance, the star 
has $M_V = \mvB$, placing it 3.5 mag above the mean main--sequence of
stars in the Solar neighborhood \citep{wright04}. Like most evolved stars,
\starB\ is chromospherically quiet with \shk~$=$~\shkB\ and
\rphk~$=$~\rhkB\ \citep{wright04b}. 
Its low chromospheric activity and location in the H--R
diagram  indicate that \starB\ is most likely a luminosity
class IV star on the subgiant branch, rather than a class V dwarf.

\starB\ is listed in the SPOCS Catalog
\citep{valenti05}  with a  metal abundance slightly below Solar
(\feh~$ = \feB \pm 0.04$) and projected rotational velocity 
\vsini~$= \vsiniB$~\ks. Interpolation of the  
star's $B-V$ color,  absolute magnitude and metallicity onto the
\citet{girardi02} stellar model grids yields a stellar mass $M_* =
\mstarB$~\msun, radius $R_* = \rstarB$~\rsun, and an age of
\ageB~Gyr. The interior models of \citet{takeda06} yield
$M_* = 1.52$~\msun and $R_* = 3.72$~\rsun. The
SPOCS Catalog lists $M_* = 
1.74$~\msun, and $R_* = 4.11$~\rsun\ \citep{valenti05}. The
variances of these different mass and 
radius estimates are 0.1~\msun\ and 0.2~\rsun,
respectively, which are consistent with our estimate of uncertainties 
in \S \ref{stellar}. The other stellar properties are listed in
Table~\ref{stellartable}. 

\begin{figure}[t!]
\epsscale{1}
\plotone{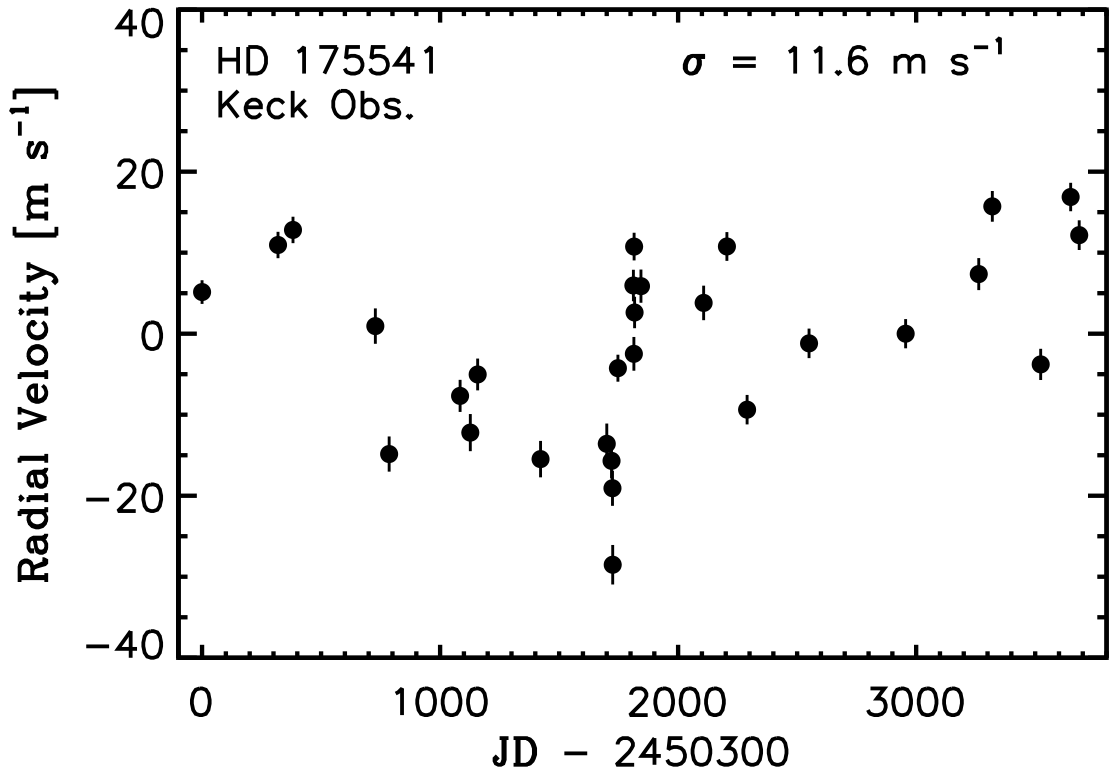}
\caption{\footnotesize{Radial velocities for \starB\ measured at Keck
  Observatory. The error bars represent the internal uncertainty of
  each measurement without accounting for stellar jitter. \label{velplotB}}   }
\end{figure}

\begin{figure}[h!]
\epsscale{1}
\plotone{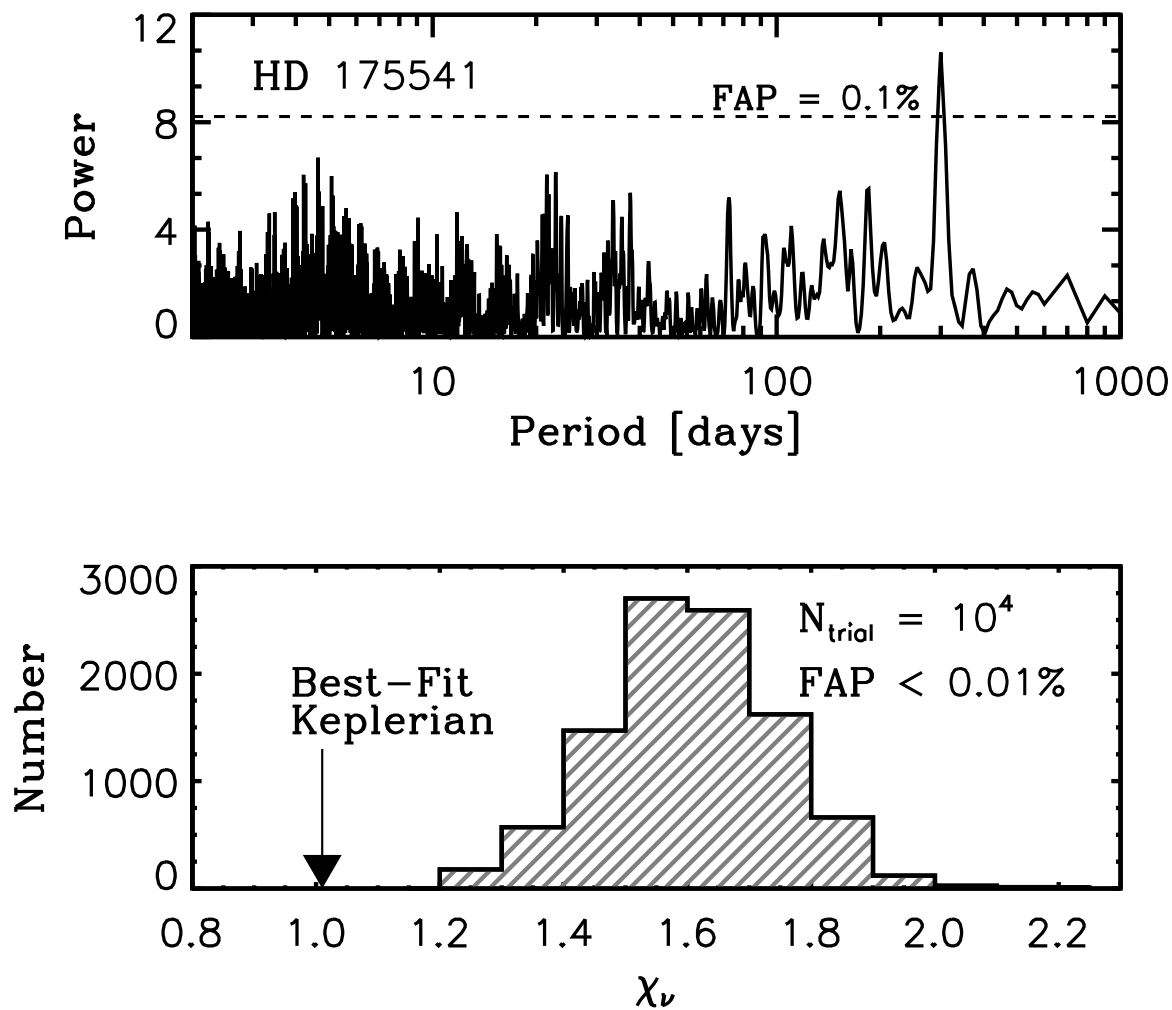}
\caption{\footnotesize{\emph{Top:} Periodogram analysis of the RV time series of
  \starB. A strong peak is visible near $P = 300$~d with an analytic
  false--alarm probability FAP$< 0.1$\%. \emph{Bottom:} Empirical assessment of
  the FAP of the best--fit Keplerian model. The original, unscrambled
  velocities yield an orbital solution with \chisq$ = \chiB$ (arrow).
  The histogram shows the distribution of \chisq\ obtained from the
  best--fit orbital solution for each of the scrambled--velocities trials.
  None of the $10^4$ trials
  produced a value of \chisq\ lower than the value obtained from the
  original time series, resulting in  FAP$< 0.01$\% (cf
  \S~\ref{sectionB}). \label{pgramB}}    }
\end{figure}
\clearpage

\begin{figure}[h!]
\epsscale{1}
\plotone{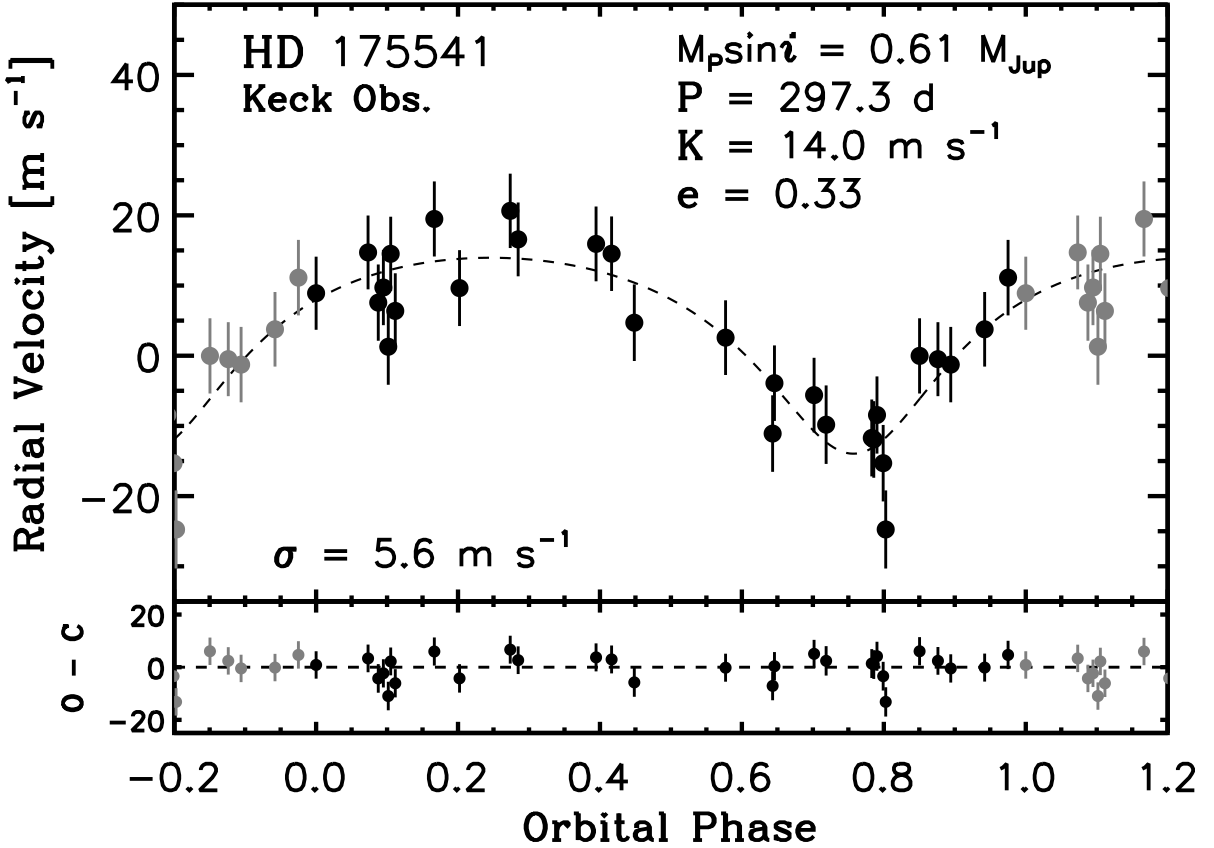}
\caption{\footnotesize{Radial velocity observations of \starB\ phased at 
  $P = \pB$~d. The gray points lie outside of phases 0.0 and 1.0 and
  are included to guide the eye. The dashed line shows the best-fit
  Keplerian orbital solution, which has \chisq$=\chiB$. \label{orbitB}}   }
\end{figure}

\starB\ was one of the original stars added to the CCPS Keck program in
1996, and was subsequently added to our list of intermediate--mass
stars in 2004. Table \ref{vel175541} lists our \nobsB\ Doppler
measurements along with their observation dates and internal
uncertainties (without jitter). Figure \ref{velplotB} shows that the rms scatter of
the velocity measurements is a factor of 6 greater than the mean internal
uncertainty ($\bar{\sigma_v} \approx 2$~\ms), and 2--3 times larger
than the rms scatter of stable Keck subgiants (Figure \ref{std_stars}). 
A Lombe--Scargle periodogram analysis of the velocities reveals a
pronounced peak near $P=300$~d, with an analytical false--alarm
probability $\rm < 0.1\%$ (Figure \ref{pgramB}).

To search for the best--fit orbital solution, we added 5~\ms\ of
jitter in quadrature to the internal measurement uncertainties. We find
that a Keplerian with $P=\pB$~d, $K = \kB$~\ms\ 
and $e = \eB$ provides the best fit to the data, resulting in
$\rm rms = \rmsB$~\ms\ and \chisq~$ = $~\chiB. Figure \ref{orbitB}
shows the the radial velocities
phased at $P = \pB$~day, along with the best--fit orbital solution
(the gray points show the measurements at phases outside of phases 0.0
and 1.0, in order to guide the eye). Assuming a stellar
mass of \mstarB~\msun, we estimate a minimum planet mass \msini~$=
\msiniB$~\mjup\ and orbital separation $a = \arelB$~AU.

While the strong periodogram peak and low \chisq\ are 
indicative of a correlated signal resulting from an orbiting planet,
it is still possible that random variability could 
conspire to produce a false periodicity in our sparse series of
measurements. To test the null hypothesis, we used the
``scrambled'' velocity false--alarm test described by
\citet{marcy05b}. For $10^4$ separate trials, we held the observation
times constant and scrambled the order of the measurements using a
pseudo random number generator. This has the effect of keeping the
sampling constant while removing any true temporal coherence, if such
a signal exists. For each
of the scrambled trials, we perform a full search for the
best--fit Keplerian orbital solution---with jitter---and record the
\chisq\ from the fit. 

The distribution of \chisq\ generated from the scrambled--velocity
trials is then compared to the fit obtained from the original time
series, as shown in the lower panel of Figure \ref{pgramB}. None of the
$10^4$ scrambled trials produced a \chisq\ equal to or lower than the
best--fit solution to the original time series, resulting in a
false--alarm probability of $< 0.01$\%. From this test, we 
conclude that the temporally correlated signal seen in the velocity
time series is likely real, rather than an artifact of random
noise. We find that the best explanation of the periodic signal is the
presence of an unseen planetary companion orbiting \starB. 

\section{Summary and Discussion}
\label{summary}

\begin{figure}[t!]
\epsscale{1}
\plotone{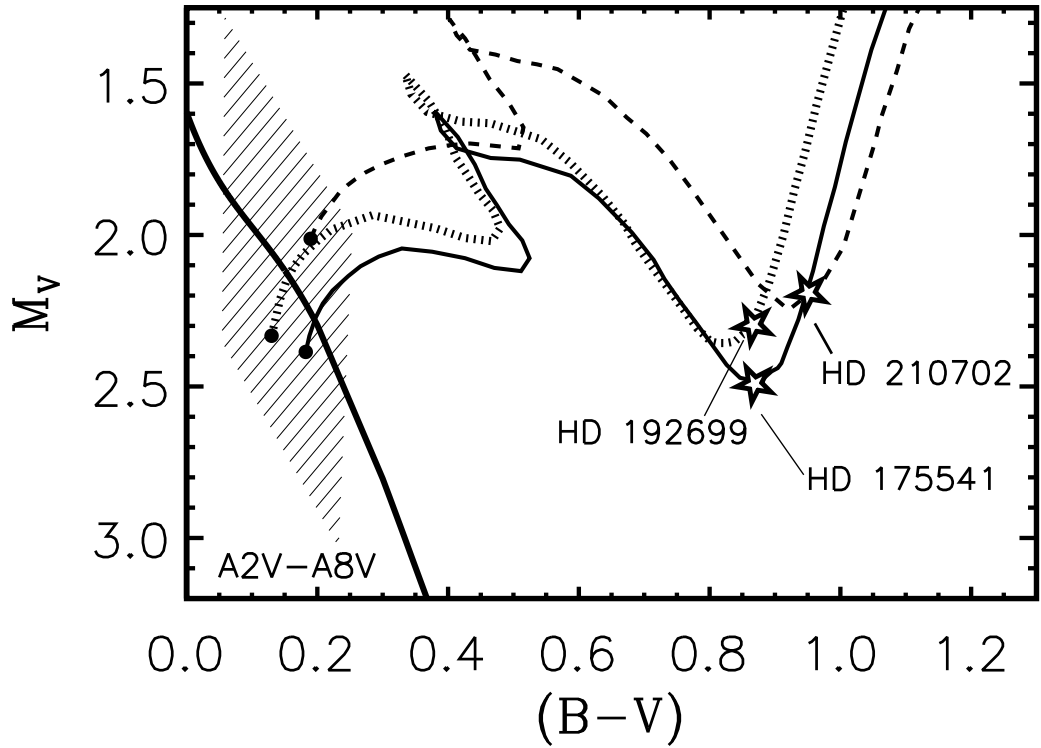}
\caption {\footnotesize{H--R diagram illustrating the properties of the
  \npllet\ subgiant planet host stars (pentagrams) compared to their
  main--sequence progenitors (filled circles). The connecting lines
  represent each star's \citet{girardi02} theoretical mass track,
  interpolated   for that star's metallicity. The thick, diagonal line
  is the  theoretical zero--age main sequence for [Fe/H]=0.0. The hashed
  region shows the approximate range of colors and magnitudes of stars with
  spectral types A2V--A8V. \label{sg_hr}}   }
\end{figure}

We present precision Doppler measurements of \npllet\
intermediate--mass subgiants that show periodic variations
in their radial velocities consistent with planet--mass orbital
companions. Interpolation of the stars' absolute magnitudes, colors
and metallicities onto the \citet{girardi02} stellar interior models
shows that all \npllet\ stars have 
masses ranging from 1.65~\msun\ to 1.85~\msun. Figure \ref{sg_hr} shows these
massive host stars on an H--R diagram, along with their theoretical
evolution tracks. Following the tracks back to the zero--age
main sequence reveals that these present--day subgiants were originally
early--type dwarfs with $B-V \lesssim 0.2$ and spectral types
ranging from A2V to A5V. The \npllet\ long--period planets presented here
would not have been detectable during their stars'
main--sequence phases due to the jitter and rotational line broadening
typical for intermediate--mass dwarfs. These planets orbiting
``retired'' A stars illustrate how evolved stars provide a unique
window into stellar mass and planetary domains otherwise
inaccessible to Doppler--based planet searches. 

There are now 9 former 
A--type stars ($1.6 \lesssim M_* < 3.0$~\msun) with planetary
companions. We list some of the properties of these massive host
stars and their planets in 
Table~\ref{massive_table}. All 9 planets orbit beyond
$\sim$0.78~AU from their stars. This paucity of planets with semimajor
axes $a < 0.78$~AU is unlikely to be due to a detection bias. For
a given planet mass and stellar mass, the velocity
semiamplitude of a star scales as $K \sim a^{-1/2}$, making planets in
smaller orbits easier to detect. The detectability
of close--in planets is also facilitated by the increased number of
orbital cycles that are observable over a given time span. 

We consider two possible explanations for the observed lack of
close--in planets around intermediate--mass stars. The first
possibility is that planets around A--type stars have the same
semimajor axis distribution as planets orbiting lower--mass stars, but
the close--in planets were destroyed by the expanding atmospheres of
their giant host 
stars. Alternatively, planets orbiting A--type stars may have a
different semimajor axis distribution than lower--mass stars, with
planets residing preferentially in long--period orbits beyond
$\sim0.8$~AU. 

These possibilities can be explored by comparing the properties of
planets in Table~\ref{massive_table} to planets orbiting lower--mass
stars listed in the CNE. We exclude extremely low--mass planets with
$K < 15$~\ms\ that would not be easily detectible around higher--mass
subgiants and giants. We use a one--sided Kolmogorov--Smirnov (K--S)
test to compare the semimajor axis distributions of  planets around
intermediate--mass and lower--mass stars \citep{press}. We find the
probability that the two distributions are identical is only
0.06\%. Under the assumption that the semimajor axis distribution of
planets is independent of stellar mass, short--period planets orbiting
evolved A--type stars must be efficiently destroyed by the expanding
atmospheres of their giant host stars. The validity of this
hypothesis depends on whether the radii of $\sim2$~\msun\
giants are large enough to engulf planets out to $\sim0.8$~AU.

\begin{figure}
\epsscale{1}
\plotone{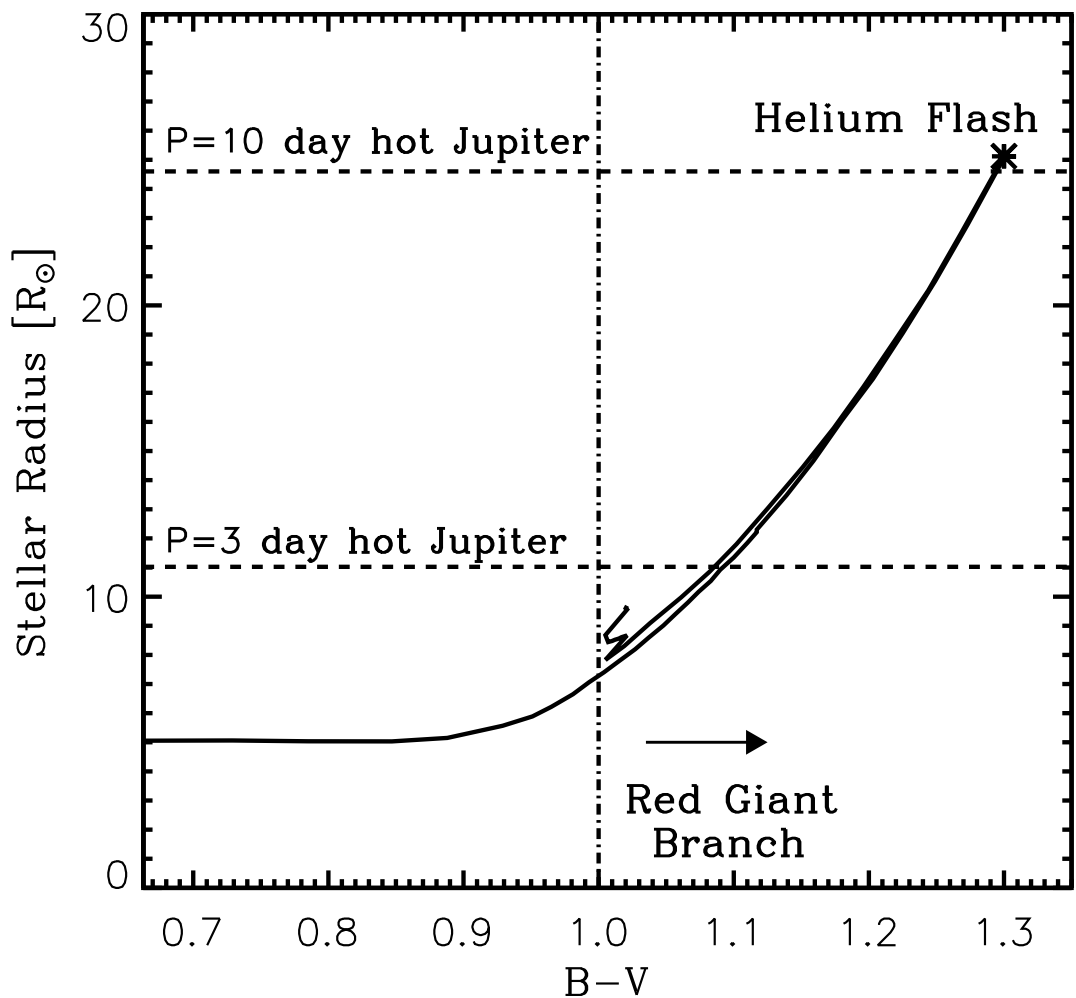}
\caption {Radius of a 1.9~\msun\ star as it evolves off of the main
  sequence (becoming redder). The horizontal dashed line s
  depict the semimajor axes of planets with periods of 3 and 10
  days. The vertical 
  dot--dashed line shows the approximate $B-V$ color of the star as it
  begins to   ascend the red giant branch. Because the radii of
  subgiants are small enough to 
  avoid interference with close--in planets, our Doppler survey
  is sensitive to the same range of orbital separations as
  surveys of main--sequence stars. Planets with $a \lesssim 30$~\rsun\ may
  be destroyed by the expansion of their host stars on the red giant
  branch. \label{rstar_hj}}  
\end{figure}

Figure~\ref{rstar_hj} shows the evolution of the radius of a
2.0~\msun\ star according to the \citet{girardi02} stellar evolution
models. As the star crosses the Hertzsprung Gap during its subgiant
phase, its radius remains nearly constant at $a = 5$~\rsun~$ =
0.023$~AU, which is within the orbit of a $P = 3$~day hot
Jupiter. Not until the star begins to ascend the RGB  
does its outer atmosphere begin to encroach on the orbits of 
short--period planets. But even at the tip of the RGB (near the helium
flash), the radius of a
2~\msun\ star is only at the distance of a 10~day hot Jupiter at $a
\approx 26$~\rsun~$=0.12$~AU (the radius of a 2.5~\msun\ red giant is
not much larger at $a \approx 32$~\rsun~$=0.15$~AU). Thus, engulfment
cannot be solely responsible for the lack of close--in
planets around subgiants and K giants. Indeed, engulfment can only be
important for 4 of the 
stars in Table~\ref{massive_table}: the post--helium--flash clump
giants HD\,104985, HD\,11977 and $\epsilon$~Tau, and HD\,13189 which
has a poorly constrained radius due to its highly uncertain parallax.  

The evolution of planetary orbits from 0.05--0.15~AU in the presense of
an expanding stellar atmosphere has not been examined in detail. The
effects of planet engulfment on its host star have been studied by
\citet{siess99}, but a key assumption in their model is that the
substellar companion is destroyed. Since it is unclear what
happens to a planet when it interacts with the atmosphere of its
expanding host star, we simply assume that planets orbiting within the radius
of a giant star are destroyed\footnote{\citet{maxted06} discovered 
a short--period substellar companion that apparently survived
engulfment as its parent star evolved into a white dwarf. However, no
Jovian planet has yet been detected around a white dwarf.}.
Under this assumption, we would expect a deficiency of hot Jupiters
around clump giants out to $\sim$0.15~AU, but no corresponding
deficiency around subgiants and K giants. 

We now analyze the lack of close--in planets around the sample in
Table~\ref{massive_table} accounting the possible destruction of hot
Jupiters around clump giants. For subgiants and giants we can use the
K--S test as before, which yields a probability of 0.7\% that the
semimajor axis distribution is the same as lower--mass stars in the
CNE. For clump giants we exclude planets from the CNE with $a <
0.15$~AU, and the corresponding  probability from the K--S test is
1.7\%.  Thus, the distribution of close--in 
planets around former A--type stars remains inconsistent with the
distribution of planets in the CNE. Since engulfment does not provide
an adequate explanation for the lack of close--in planets in
Table~\ref{massive_table}, we are left  with the possibility that the
semimajor axis distribution of planet around A--type stars is
significantly different than the distribution around lower--mass stars
($M_* < 1.6$~\msun).    

Differences between the semimajor axes of planets around stars of
various masses has previously been investigated by \citet{burkert06}.
From their study of the orbital properties of known exoplanets, they
find evidence of a gap in the semimajor axis distribution 
around stars with masses $M_* \geq 1.2$~\msun, with fewer planets
between 0.08 AU and 0.6 AU compared to lower--mass stars. They were
able to reproduce this gap in their Monte Carlo simulations of planet 
migration, and they attribute the gap to the shorter depletion
timescales of disks around intermediate--mass stars.

The semimajor axis distribution of planets as a function of stellar
mass can be investigated further with the inclusion of a larger sample
of intermediate--mass subgiants in Doppler-based planet searches. As
Figure~\ref{rstar_hj} shows, Doppler surveys of subgiants can probe 
occurrence of Jovian planets at orbital distances ranging from many AU
down to as close as 0.05~AU, the realm of hot Jupiters. The smaller radii of
subgiants also result in  higher surface gravities compared to giants,
which leads to lower levels of pulsation--induced
jitter. \citet{hekker06} show that giants  with $B-V > 1.2$ typically
have jitter values greater than 20~\ms, ostensibly due to radial and
non--radial pulsation modes. Only giants blueward of this limit are
stable to within 20~\ms, compared to the 4--6~\ms\ of jitter seen in
subgiants ($B-V < 1.0, M_V \lesssim 2.0$). This increased velocity
stability, coupled with their relatively small radii, therefore make
subgiants ideal proxies for A--type dwarfs in Doppler--based planet
searches.   

The primary limitation of subgiants is their relative scarcity, which
restricts the number of bright targets suitable for high--resolution
spectroscopic observations. The
time it takes stars to cross the Hertzsprung Gap is small compared to
the star's lifetime---of order 100~Myr---rendering Hertzsprung Gap
stars within 200~pc rare compared to main--sequence stars and
giants. Additional targets can be 
found further from the Sun, with fainter apparent magnitudes ($V
\gtrsim 7.5$). In the near future, we plan to expand our sample of
subgiants using the Keck telescope and HIRES spectrometer in order to
further investigate the orbital properties, planet masses and
occurrence rate of planets orbiting intermediate--mass stars. 
As the number of subgiants included in
Doppler surveys increases, it will become apparent whether the lack
of short--period planets around intermediate--mass stars is a result of
different formation and migration mechanisms in the disks of A--type
stars, or simply a consequence of the small number of massive subgiants
currently surveyed. 

\acknowledgements 

We extend our gratitude to the many CAT observers who have helped 
with this project, including Howard Isaacson, Julia Kregenow, Karin
Sandstrom, Bernie 
Walp, Peter Williams, Katie Peek and Shannon Patel. Special thanks 
to Herv\'e Buoy and Francisco Ramos-Stierle for lending a portion of
their 3m observing time to observe HD\,192699 before it set in 2006. 
We thank Michael Fitzgerald and Marshall Perrin for their useful
discussions, and Tim Robishaw for sharing his data display expertise
and IDL plotting routines. We 
also gratefully acknowledge the efforts and dedication of the Lick
Observatory and Keck Observatory staff, and the time assignment
committees of NASA, NOAO and University of California for their
generous allocations of observing time. We appreciate funding from
NASA grant NNG05GK92G (to GWM), and the NSF for its grant AST-0307493
(to SSV) for supporting this research. DAF is a Cottrell 
Science Scholar of Research Corporation and acknowledges support from
NASA Grant NNG05G164G that made this work possible. This research has
made use of the Simbad database operated at CDS, Strasbourg France,
and the NASA ADS database. The authors wish to
extend special thanks to those of Hawaiian ancestry on whose sacred
mountain of Mauna Kea we are privileged to be guests. Without their
generous hospitality, the Keck observations presented herein would not
have been possible.

\begin{deluxetable}{lcllll}
\tablecaption{Planet Host Stars With $M_* \geq 1.6$~\msun\tablenotemark{a} \label{massive_table}}
\tablewidth{0pt}
\tablehead{\colhead{HD}  & 
\colhead{Spectral Type}  &
\colhead{$M_*$ (\msun)} &
\colhead{$R_*$ (\rsun)} &
\colhead{$a$ (AU)} &
\colhead{References} \\
}
\startdata
13189   & K2\,II  & 2--6\tablenotemark{b}  &  ...  & 1.5--2.2  & 1\\
28305\tablenotemark{c}  & K0\,III & 2.7   &  13.7 & 1.93 & 2\\  
11977   & G5\,III & 1.9       & 13 & 1.93 & 3\\
62509\tablenotemark{d} & K0\,III & 1.86  & 8.8 & 1.7  & 4,5\\  
210702  & K1\,IV  & \mstarA   & \rstarA & \arelA & 6  \\
192699  & G8\,IV  & \mstarC   & \rstarC & \arelC & 6  \\
175541  & G8\,IV  & \mstarB   & \rstarB & \arelB & 6  \\
89744   & F7\,IV  & 1.65      & 1.1 & 0.93  & 7 \\
104985  & G9\,III   & 1.6     & 8.9 & 0.78 & 8 \\
\enddata
\tablenotetext{a}{Excluded from this list of evolved, 
intermediate--mass planet host stars are
$\gamma$\,Cep~A and HD\,47536. The discovery paper  by
\citet{hatzes03} cites a  
  stellar mass of 1.59~\msun. However, \citet{torres07} find a much
  lower value of $M_* =   1.18^{+0.04}_{-0.11}$~\msun\ by using a
  theoretical model grid interpolation, and a recent dynamical
  analysis by  \citet{neu07} yields a stellar mass of $1.40 \pm
  0.12$~\msun. For the  clump giant HD\,47536, \citet{setiawan03}
  list a highly  uncertain mass with a lower limit of 1.1~\msun, which
  falls well below the 1.6~\msun\ cutoff used for this table. Also omitted
  are the 3 intermediate--mass K giants in
  \citet{mitchell04}, which have not yet been published in a refereed
  journal.}   
\tablenotetext{b}{\citet{schuler05}} 
\tablenotetext{c}{HD\,28305~$ =\epsilon$~Tau.} 
\tablenotetext{d}{HD\,62509~$ = \beta$~Gem~$ = $~Pollux.} 
\tablerefs{(1) \citet{hatzes05}; (2) \citet{sato07}; (3)
  \citet{setiawan05}; (4) \citet{hatzes06}; (5) \citet{reffert06}; (6)
this work; (7) \citet{korzennik00}; (8) \citet{sato03}}
\end{deluxetable}

\clearpage
\begin{deluxetable}{lllll}
\tablecaption{Stellar Parameters\label{stellartable}}
\tablewidth{0pt}
\tablehead{\colhead{Parameter}  & 
\colhead{\starC} & 
\colhead{\starA} &
\colhead{\starB} \\
}
\startdata
V              & \vmagC          & \vmagA         & \vmagB         \\
$M_V$          & \mvC            & \mvA           & \mvB           \\
B-V            & \bvC            & \bvA           & \bvB           \\
Distance (pc)  & \dC             & \dA            & \dB            \\
${\rm [Fe/H]}$ & \feC~(0.04)     & \feA~(0.04)    & \feB~(0.04)    \\
$T_{eff}$~(K)  & \teffC~(44)     & \teffA~(44)    & \teffB~(44)    \\
\vsini~(\ks)   & \vsiniC~(0.5)   & \vsiniA~(0.5)  & \vsiniB~(0.5)  \\
$\log{g}$      & \loggC~(0.3)    & \loggA~(0.3)   & \loggB~(0.3)   \\
$M_{*}$~(\msun) & \mstarC~(0.12) & \mstarA~(0.13) & \mstarB~(0.12) \\
$R_{*}$~(\rsun) & \rstarC~(0.51) & \rstarA~(0.57) & \rstarB~(0.46) \\
$L_{*}$~(\lsun) & \lstarC~(0.3)  & \lstarA~(0.3)  & \lstarB~(0.3)  \\
Age~(Gyr)      & \ageC~(1.0)     & \ageA~(1.0)    & \ageB~(1.0)    \\
$S_{HK}$       & \shkC           & \shkA          & \shkB          \\
$\log R'_{HK}$ & \rhkC           & \rhkA          & \rhkB          \\
\enddata
\end{deluxetable}

\begin{deluxetable}{lll}
\tablecaption{Radial Velocities for HD~192699\label{vel192699}}
\tablewidth{0pt}
\tablehead{
\colhead{JD} &
\colhead{RV} &
\colhead{Uncertainty} \\
\colhead{-2440000} &
\colhead{(m~s$^{-1}$)} &
\colhead{(m~s$^{-1}$)} 
}
\startdata
13155.988 &  -60.66 & 5.14 \\
13257.741 &    3.44 & 6.31 \\
13522.921 &  -47.84 & 4.90 \\
13576.901 &  -20.91 & 5.33 \\
13619.774 &   13.49 & 4.90 \\
13640.713 &   26.10 & 4.99 \\
13668.641 &   19.02 & 4.66 \\
13879.905 &  -46.41 & 4.39 \\
13891.854 &  -34.19 & 5.46 \\
13894.944 &  -42.03 & 4.94 \\
13895.904 &  -56.83 & 4.71 \\
13896.937 &  -50.29 & 4.57 \\
13905.962 &  -28.97 & 6.96 \\
13921.932 &  -18.30 & 5.37 \\
13922.851 &   -1.89 & 4.63 \\
13926.856 &  -18.51 & 4.47 \\
13951.801 &   11.70 & 4.62 \\
13959.771 &   -4.84 & 5.15 \\
13966.784 &    8.99 & 5.63 \\
13975.738 &    5.99 & 4.65 \\
13976.714 &   14.68 & 4.87 \\
13998.725 &   25.11 & 5.28 \\
14001.749 &   34.99 & 4.68 \\
14020.654 &   49.36 & 5.02 \\
14021.682 &   25.41 & 5.42 \\
14034.638 &   18.33 & 5.23 \\
14035.671 &    9.67 & 5.46 \\
14039.603 &   -2.60 & 5.69 \\
14046.658 &    6.44 & 5.05 \\
14049.615 &    5.84 & 5.60 \\
14059.610 &   -5.29 & 4.55 \\
14070.664 &  -16.58 & 5.30 \\
14072.612 &  -18.70 & 5.17 \\
14170.058 &  -78.23 & 5.92 \\
\enddata
\end{deluxetable}

\clearpage
\begin{deluxetable}{lllll}
\tablecaption{Orbital Parameters\label{orbittable}}
\tablewidth{0pt}
\tablehead{\colhead{Parameter}  & 
\colhead{\starC\,b} & 
\colhead{\starA\,b} &
\colhead{\starB\,b} \\
}
\startdata
P (d)              & \pC~(\peC)     & \pA~(\peA)    & \pB~(\peB)      \\
T$_p$\tablenotemark{a}~(JD) & \tpC~(\tpeC) & \tpA~(\tpeA) & \tpB~(\tpeB)  \\
e                  & \eC~(\eeC)     & \eA~(\eeA)    & \eB~(\eeB)      \\
K$_1$~(\ms)        & \kC~(\keC)     & \kA~(\keA)    & \kB~(\keB)      \\
$\omega$~(deg)     & \omC~(\omeC)   & \omA~(\omeA)  & \omB~(\omeB)    \\
\msini~(\mjup)     & \msiniC        & \msiniA       & \msiniB         \\
$a$~(AU)           & \arelC         & \arelA        & \arelB          \\
Fit RMS~(\ms)      & \rmsC          & \rmsA         & \rmsB           \\
\chisq             & \chiC          & \chiA         & \chiB           \\
N$_{\rm obs}$      & \nobsC         & \nobsA        & \nobsB          \\
\enddata
\tablenotetext{a}{Time of periastron passage.}
\end{deluxetable}

\begin{deluxetable}{lll}
\tablecaption{Radial Velocities for HD~175541\label{vel175541}}
\tablewidth{0pt}
\tablehead{
\colhead{JD} &
\colhead{RV} &
\colhead{Uncertainty} \\
\colhead{-2440000} &
\colhead{(m~s$^{-1}$)} &
\colhead{(m~s$^{-1}$)} 
}
\startdata
10283.952 &    5.14 & 1.46 \\
10603.027 &   10.95 & 1.63 \\
10665.908 &   12.81 & 1.63 \\
11011.878 &    0.94 & 2.19 \\
11069.830 &  -14.86 & 2.17 \\
11367.894 &   -7.68 & 1.98 \\
11410.857 &  -12.21 & 2.29 \\
11441.759 &   -5.03 & 1.96 \\
11705.979 &  -15.48 & 2.25 \\
11984.164 &  -13.59 & 2.50 \\
12004.126 &  -15.69 & 2.18 \\
12008.065 &  -19.07 & 2.17 \\
12009.118 &  -28.52 & 2.44 \\
12030.989 &   -4.26 & 1.67 \\
12095.962 &    5.95 & 1.94 \\
12098.016 &   -2.49 & 2.09 \\
12099.039 &   10.75 & 1.71 \\
12100.956 &    2.62 & 1.94 \\
12127.874 &    5.87 & 2.05 \\
12391.132 &    3.80 & 2.13 \\
12488.898 &   10.77 & 1.78 \\
12573.725 &   -9.38 & 1.82 \\
12833.948 &   -1.19 & 1.82 \\
13239.787 &    0.00 & 1.80 \\
13546.914 &    7.36 & 1.98 \\
13603.852 &   15.72 & 1.89 \\
13807.150 &   -3.79 & 1.93 \\
13932.962 &   16.87 & 1.76 \\
13968.920 &   12.16 & 1.83 \\
\enddata
\end{deluxetable}

\begin{deluxetable}{lll}
\tablecaption{Radial Velocities for HD~210702\label{vel210702}}
\tablewidth{0pt}
\tablehead{
\colhead{JD} &
\colhead{RV} &
\colhead{Uncertainty} \\
\colhead{-2440000} &
\colhead{(m~s$^{-1}$)} &
\colhead{(m~s$^{-1}$)} 
}
\startdata
13241.863 &   18.13 & 3.32 \\
13256.776 &    9.30 & 3.91 \\
13310.752 &  -16.07 & 4.33 \\
13342.598 &   -0.18 & 3.37 \\
13569.930 &   42.99 & 4.77 \\
13573.904 &   29.79 & 5.13 \\
13574.927 &   38.71 & 4.47 \\
13575.897 &   29.80 & 4.41 \\
13640.779 &   10.99 & 5.45 \\
13922.930 &   28.01 & 3.39 \\
13959.860 &    0.18 & 3.22 \\
13975.798 &   -1.47 & 3.54 \\
13976.764 &   -8.51 & 3.20 \\
13998.768 &  -10.73 & 3.52 \\
14001.789 &  -13.09 & 3.61 \\
14020.721 &  -15.97 & 3.73 \\
14021.735 &  -11.53 & 4.38 \\
14034.695 &  -22.55 & 4.41 \\
14046.694 &  -31.56 & 3.40 \\
14059.673 &  -14.62 & 3.57 \\
14070.705 &  -11.10 & 3.70 \\
14092.594 &    1.07 & 4.06 \\
14097.649 &   -2.40 & 3.31 \\
14098.603 &    1.90 & 3.78 \\
14103.598 &    9.35 & 3.65 \\
14109.626 &    9.02 & 5.12 \\
14117.595 &   22.46 & 5.58 \\
14136.600 &   38.79 & 3.09 \\
14197.033 &   53.95 & 4.26 \\
\enddata
\end{deluxetable}

\end{document}